\documentstyle[11pt,paspconf,epsf]{article}
\begin{document}

\title{The Stellar Mass Function}

\author{Pavel Kroupa\footnotemark[1]} \footnotetext[1]{e-mail:
pavel@ita.uni-heidelberg.de; To appear in PASPC: {\it Brown Dwarfs and
Extrasolar Planets}, R. Rebolo, M.R. Zapatero Osorio \& E. Martin
(eds.); International workshop held on Tenerife Island,
March~17~--~21, 1997.}

\affil{Institut f{\"u}r Theoretische Astrophysik, Universit{\"a}t
Heidelberg, Tiergartenstr. 15, D-69121 Heidelberg, Germany}

\begin{abstract}
Since the major review by Scalo (1986), significant progress has been
achieved in constraining the mass function (MF) of low-mass stars.
The break-throughs which today allow a much better understanding of
the stellar luminosity function (LF) and the underlying MF are
documented here and the resulting MF for Galactic field stars is
confronted with microlensing data, the chemical enrichment history of
the Galaxy, the Oort limit and well-studied open clusters.
\end{abstract}


\section{Introduction}
The mass, $m$, of an isolated main-sequence star can be determined
from its absolute luminosity, $l$, and the mass--luminosity relation,
$m(l, {\rm [Fe/H]}, \tau,\mbox{\boldmath$s$})$.  The corresponding
mass--(absolute-magnitude) relation is $m(M_P)$, where $P$ represents
some photometric pass band, and the other parameters have been dropped
for conciseness. For Galactic disc stars with $m<1\,M_\odot$, the
stellar age is usually $\tau>1$~Gyr, [Fe/H]$>-1$ and the dependence on
the stellar spin, {\boldmath$s$}, is not significant.  The continuous
distribution of stars by luminosity is related to the distribution by
mass through $\psi(M_P) = - \xi(m) \times (dm/dM_P)$, where
$\psi(M_P)$ and $\xi(m)$ are the LF and MF, respectively.  The
differential volume number density of stars in the absolute-magnitude
interval $M_P+dM_P$ to $M_P$ is $d\rho=-\psi\,dM_P$, which is also the
differential number density of stars, $\xi\,dm$, in the mass range $m$
to $m+dm$.

The conversion of an observed LF to the MF is demonstrated in Fig.~1
using two well-cited examples from the literature.  The MF is a good
approximation to the initial mass function (IMF) for stars with
$M_V>5$, because they do not evolve appreciably within the age of the
Galactic disc.

\begin{figure} 
\plotone{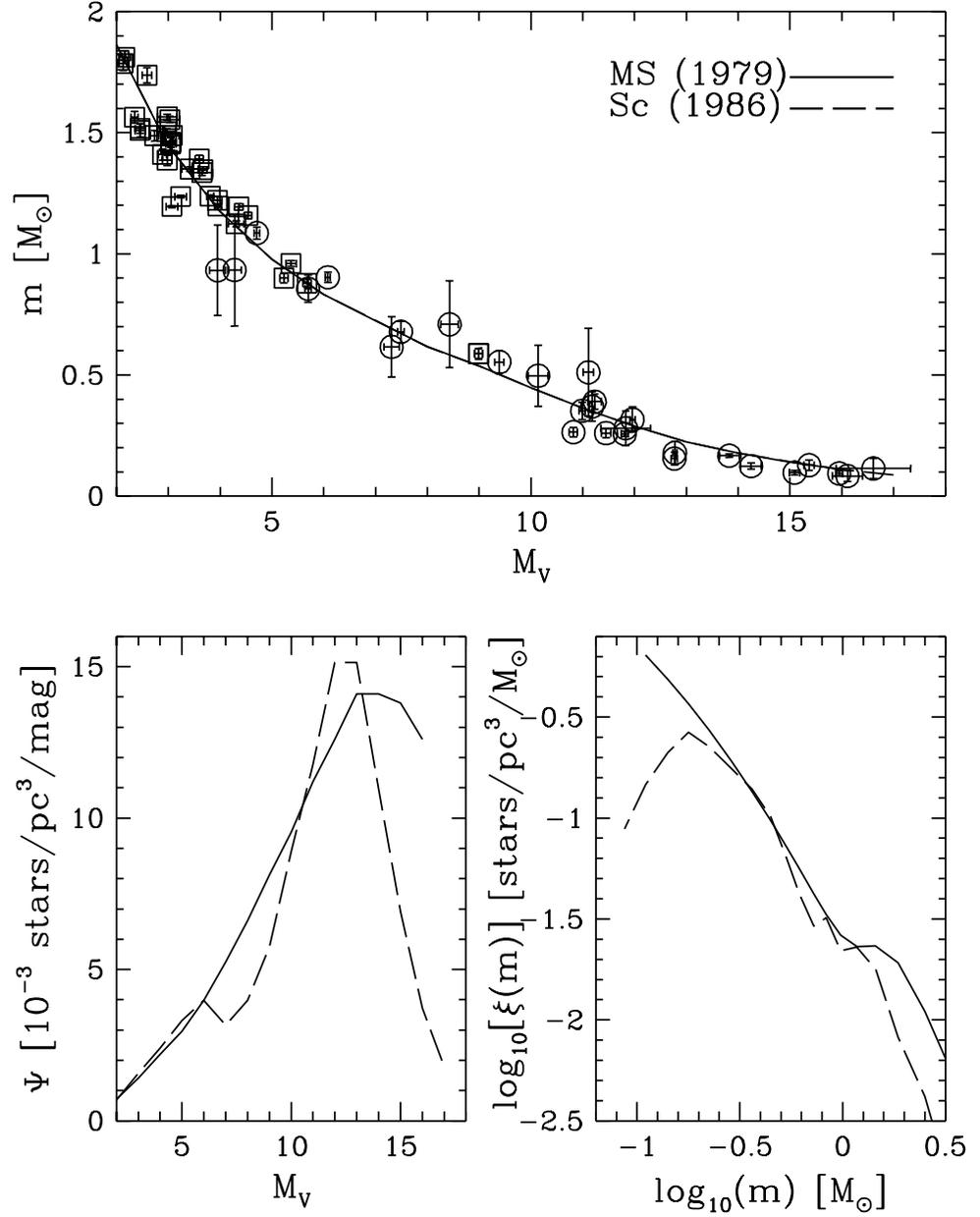}
\caption{Upper panel: The solid and indistinguishable dashed line are
the $m(M_V)$ relations used by MS and Sc, respectively. Open circles
are observational data from HM93, and the open squares are from
Andersen (1991). Lower panels: The stellar LF adopted by MS (solid
line) and by Sc (dashed line) is shown in the left panel, and the
resulting IMFs are shown in the right panel.}
\end{figure}

The mass-$M_V$ data compiled by Henry \& McCarthy (1993, hereinafter
HM93) suggest that the $m(M_V)$ relations adopted by Miller \& Scalo
(1979, hereinafter MS) and Scalo (1986, hereinafter Sc) assign masses
that may be somewhat too small for stars with $M_V\approx10$.  As
shown in Section~4, theoretical work on the structure of low-mass
stars confirms that there is additional structure in the $m(M_V)$
relation near this magnitude.  This is an important point (D'Antona \&
Mazzitelli 1983, hereinafter D'AM83) because $\psi\propto dm/dM_V$,
and thus any mis-understood structure in the $m(M_V)$ relation carries
through to unphysical structure in the MF.  The slope, $dm/dM_V$,
becomes small at the faintest magnitudes and is not very well
determined. This holds true in other photometric passbands as well
(HM93) and leads to large uncertainties in the MF near the hydrogen
burning mass limit.

In Fig.~1 the IMF, $\xi(m)$, is plotted as the volume number density
per unit mass interval. This is obtained from the IMF, $\xi'({\rm
log}_{10}[m])$, expressed as the surface number density per unit
logarithmic mass interval. It is is given by Eqn.~30 in MS for 
a constant star formation rate and a Galactic disc age $T_0=12$~Gyr
(their Table~7), and by Sc in his Table~VII for $T_0=12$~Gyr. The
conversion is $\xi(m) = \xi'({\rm log}_{10}[m])/(2\,h\,m\,{\rm
ln}(10))$, where $2h$ is the Galactic disc thickness.

The Miller \& Scalo IMF, $\xi_{\rm MS}$, continues to increase down to
$m\approx 0.1\,M_\odot$. The Scalo IMF, $\xi_{\rm Sc}$, on the other
hand, has a significant maximum at $m\approx 0.2\,M_\odot$.  The
reason for the difference is the adopted LF, measurement of which has
been improved significantly between 1979 and~1986 through star-count
surveys in the solar neighbourhood (Wielen, Jahreiss \& Kr{\"u}ger
1983) and on deep photographic plates (Reid \& Gilmore 1982, hereinafter
RG).

Both $\xi_{\rm MS}$ and $\xi_{\rm Sc}$ have a dip near $m=0.6-1.3
\,M_\odot$. This results because star-count data have to be corrected
for the loss of the more massive stars through stellar evolution, and
because $h$ is smaller for the more massive stars. The dip can be
minimised by choosing an appropriate age for the Galactic disc
($T_0\approx 12$~Gyr) and star-formation history.  For stars brighter
than $M_V\approx6$, the Galactic disc thickness decreases from
$h\approx300$~pc to $h\approx90$~pc (e.g. Fig.~10 in Sc) leading to an
enhancement of the volume density of brighter stars near the Galactic
plane.

An additional reason for the dip in $\xi_{\rm Sc}$ is that the LF
adopted by Sc shows such a feature at $M_V\approx6-9$. This is
recognised to be real (Upgren \& Armandroff 1981) and a result of
stellar physics (D'AM83; Kroupa, Tout \& Gilmore 1990, hereinafter
KTG90).

The intricate details of the transformation of an observed single-star
LF to the IMF, taking account of the varying $h$, the birth-rate
history of stars and stellar evolution, can be found in Sc and
Haywood, Robin \& Cr\'ez\'e (1997).  Concerning low-mass stars, we
know today that (i) the smooth $m(M_V)$ relation used by MS and Sc is
not consistent with stellar models, and that (ii) the LFs adopted by
MS and Sc do not correctly represent the distribution of single stars
with luminosity. Consequently, the MFs plotted in Fig.~1 do not
correctly describe the spectrum of stellar masses.

Since~1986, constraints on the LF have been improved (Section~2), the
understanding of Malmquist bias has advanced considerably (Section~3),
the observational and theoretical constraints on the mass--luminosity
relation have been improved (Section~4), and the characteristics of
the binary star population in the Galactic field is much better
quantified (Section~5).  Discussions of the MF in the substellar
regime can be found in Laughlin \& Bodenheimer (1993), Basri \& Marcy
(1997), Chabrier (these proceedings), and Jones (these proceedings).
A review of the faint end of the LF can also be found in Bessell \&
Stringfellow (1993). 

\section{The Luminosity Function}
Trigonometric parallaxes are used to infer the LF for the stars in the
immediate neighbourhood of the Sun, where most multiple-star systems
have been identified.  Completeness extends, for northern declinations,
to~20~pc for stars with $M_V<9.5$, but severe incompleteness sets in
beyond 5~pc for stars with $M_V>12$ (Jahreiss 1994, Henry et
al. 1997).  Estimates of the nearby LF, $\psi_{\rm near}$, using
somewhat different sampling volumes, are provided by Wielen et
al. (1983), Dahn, Liebert \& Harrington (1986, hereinafter DLH) and
KTG93 in the V-band, and by HM90 in the K-band.  Three additional
stars have been found recently within 5.3~pc: Gl~866C with
$M_V\approx15.8$ (Henry, private communication; see also Simons, Henry
\& Kirkpatrick 1996 and Kroupa 1995a), LP~944--20 with $M_V>19$
(Kirkpatrick, Henry \& Irwin 1997), and LHS~1565 with $M_V\approx15.2$
(Henry et al. 1997).  However, $\psi_{\rm near}$ remains poorly
constrained for $M_V>12$, and the contribution by faint main-sequence
stars to the mass of the Galactic disc remains unclear.

Other techniques have been developed with the aim of improving the
constraints on the LF at the faint end.  RG made deep photographic
exposures with a Schmidt telescope (see also Hawkins, these
proceedings) to probe the distribution of faint main sequence stars to
distances of about 100~pc.  They used an automatic plate measuring
machine to identify Galactic-disc red-dwarf stars from the hundred
thousand images recorded on a typical Schmidt-telescope plate in the
direction of the South Galactic Pole. At least two photometric pass
bands are necessary to define the colours of the objects, and multiple
exposures on different plates are used to confirm images. The sample
of possible red dwarf stars is contaminated by red giants for
typically $V-I<1.8$, and galaxies dominate the number counts at the
faintest magnitudes ($V>19$). Distances to the stars in the final
red-dwarf sample are estimated using photometric parallax. The stellar
number densities are estimated taking into account the distance to
which a star can be detected.

After the pioneering work by RG, other deep surveys with the method of
photometric parallax explored the distribution of faint stars in
different directions, Gilmore, Reid \& Hewett (1985), Hawkins \&
Bessell (1988), Leggett \& Hawkins (1988), Stobie, Ishida \& Peacock
(1989, hereinafter SIP) and Kirkpatrick et al. (1994).  These yield
{\it photometric} LFs, $\psi_{\rm phot}$, that are consistent with
each other and, after correction for Galactic disc structure, are
independent of the direction of the line-of-sight through the Galactic
disc.  A consistent finding among these surveys, each yielding about
30--60~stars with $M_V>12$, is that $\psi_{\rm phot}$ has a maximum at
$M_V\approx12$ with a decline at fainter magnitudes.  That it also
does not show significant variation with direction has been verified
by the extensive photographic survey of Tinney, Reid \& Mould
(1993). The resulting LF is based on about 3500~stars spread over an
area of 270~square degrees.  Surveys with the Hubble Space Telescope
(HST) avoid contamination by galaxies down to $V\approx21-26$, and can
thus probe the stellar distribution to distances of a few
kpc. Santiago, Gilmore \& Elson (1996) report $\psi_{\rm phot,HST}$
based on about 20~fields and verify that $\psi_{\rm phot}$ decreases
for $M_V>12$. They discuss the uncertainties in $\psi_{\rm phot,HST}$
owing to photometric calibration.

Finally, Jarrett, Dickman \& Herbst (1994) constrain the faint star LF
by counting 14~red stars with $M_V>12$ that appear projected against
dark molecular clouds.  Significant conclusions from a comparison with
$\psi_{\rm near}$ or $\psi_{\rm phot}$ are not possible but the sample
is also interesting because it may contain young M~dwarfs that have
migrated from the molecular cloud.

\section{Malmquist Bias}
In a colour--magnitude diagram there is scatter about the average
relation, because the luminosity of a main-sequence star is a function
of mass, metallicity, age and spin (Section~1). Consequently, the
distance estimated from the colour and apparent magnitude of a star is
subject to error. In a magnitude limited survey, such as is used to
construct $\psi_{\rm phot}$, this leads to overestimation of the
stellar number densities because the number of stars increases with
distance.  More intrinsically brighter stars are counted in the
photometrically defined sampling volume than intrinsically fainter
stars are lost from it. This bias also leads to the resulting stellar
sample appearing brighter than is typical for stars of the given
colour.

SIP applied for the first time the complete Malmquist corrections to
the raw photometric LF. The Malmquist correction, $\Delta\psi_{\rm
phot}/\psi_{\rm phot}\propto \sigma_{M_P}^2$, where $\sigma_{M_P}$ is
the scatter about the mean absolute magnitude $M_P$ in the
colour--magnitude diagram, and is assumed to be constant and Gaussian.
The cosmic scatter, $\sigma_{M_P}$, depends on the photometric bands
used for photometric parallax. Using $M_I(R-I)$, rather than the
shallower $M_V(V-I)$ relation, leads to significantly larger Malmquist
bias (Reid 1991; Fig.~2 in Kroupa 1995a).

The detailed study of cosmic scatter in Section~3 of KTG93 shows that
$\sigma_{M_V}$ is non-Gaussian and a function of $V-I$, because the
effects of metallicity, unresolved binaries and age lead to different,
colour dependent changes in $M_V$.  Recent progress in the modelling
of realistic stellar atmospheres (Fig.~1 in Baraffe et al. 1995)
qualitatively supports the metallicity model of KTG93 (their Figs.~5
and~6).  Malmquist corrections are thus even more involved than the
treatment by SIP.

\section{The Mass--(Absolute-Magnitude) Relation}
The $m(M_V)$ relation steepens near $M_V=10$ and flattens again near
$M_V=15$, because the formation of H$_2$ in the outer shells of 
main-sequence stars causes core contraction. This leads to brighter
luminosities and full convection for $m<0.35\,M_\odot$ (Chabrier \&
Baraffe 1997; Kroupa \& Tout 1997, hereinafter KT), and to a pronounced
local minimum in $dm/dM_V$ at $M_V\approx11.5$. Artificial suppression
of H$_2$ formation leads to full convection at $m<0.25\,M_\odot$ and
to fainter stars (KTG90, KT). In this case $dm/dM_V$ has no
significant extrema for $M_V>9$.

\begin{figure} 
\plotone{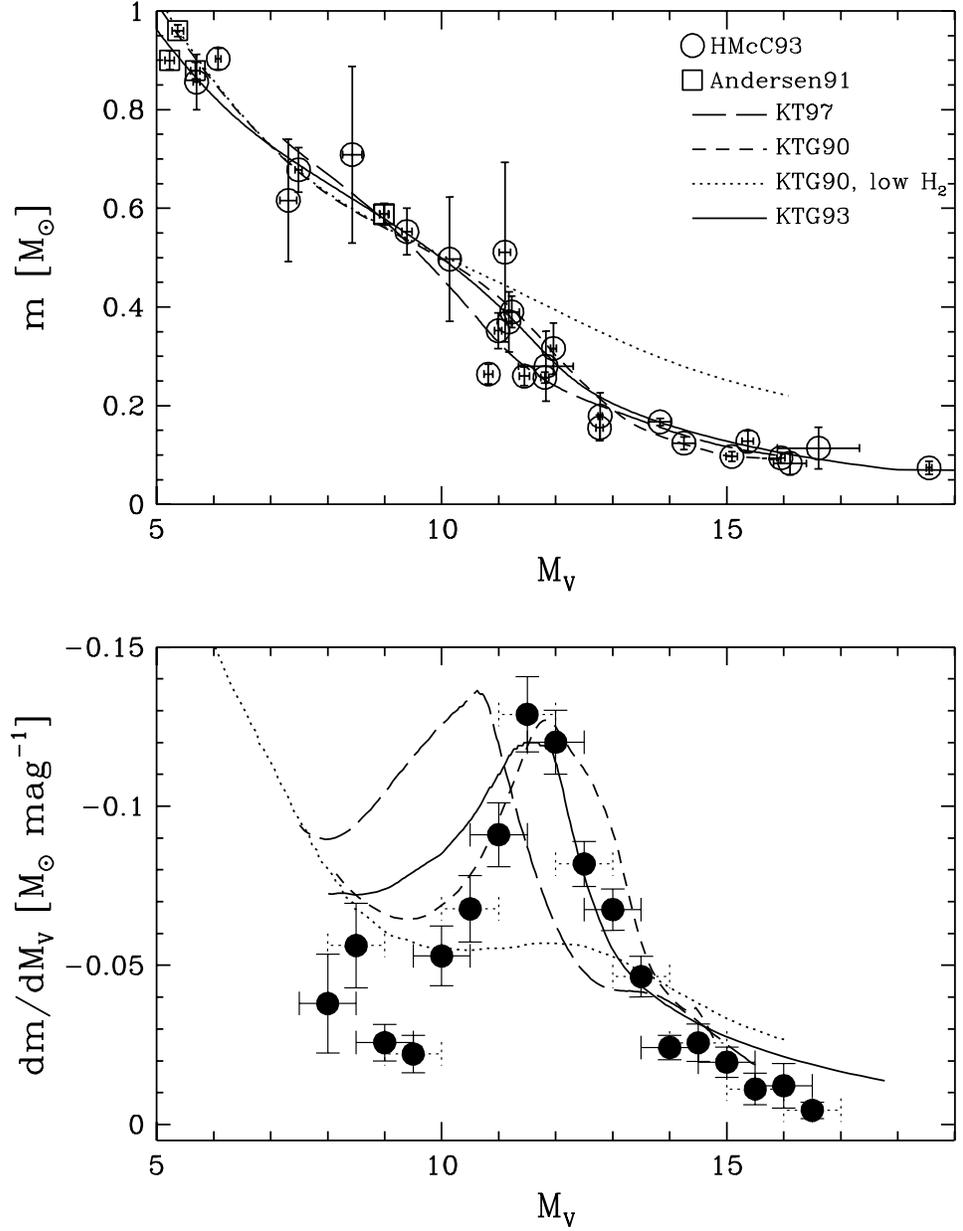}
\caption{Upper panel: Open circles and squares are observational data
as in Fig.~1.  The short-dashed and dotted lines are theoretical
$m(M_V)$ relations for normal and significantly suppressed H$_2$
abundance, respectively. More recent models are shown as the
long-dashed line.  The solid line is the empirical relation.  Lower
panel: $dm/dM_V$ for the $m(M_V)$ relations shown in the upper
panel. The solid dots are $\overline{\psi}_{\rm phot}$ (Section~6)
scaled to fit the figure. }
\end{figure}

This is illustrated in Fig.~2. The discussion in Section~1 implies
that a minimum in $dm/dM_V$ should correspond to a maximum in the LF,
unless the MF has a minimum that cancels out the structure in
$dm/dM_V$.  All photometric LFs show a maximum near $M_V=12$
(Section~2). The well constrained average photometric LF,
$\overline{\psi}_{\rm phot}$ (Section~6), shows this maximum to be in
beautiful agreement with the theoretical $dm/dM_V(M_V)$ relation,
opening the possibility of testing stellar structure theory near the
critical $m\approx0.35\,M_\odot$ where stars become fully convective.
For example, the modern stellar models of KT, that rely on the
Eddington approximation for the surface boundary condition, place the
minimum in $dm/dM_V$ at too bright an $M_V$ (long-dashed curve in
Fig.~2). Treatment of a realistic stellar atmosphere together with the
most recent stellar structure physics by Baraffe et al. (1995), leads
to improved agreement of the extremum in $dm/dM_V$ with
$\overline{\psi}_{\rm phot}$ (KT). Agreement, however, does not
necessarily imply correct physics, as the out-dated models of KTG90
(short-dashed curve in Fig.~2) demonstrates. Theoreticians must, in
addition, seek agreement with luminosity--(effective-temperature),
colour--magnitude and mass--(absolute-magnitude) relations (KT).

\section{Binary Stars}
Duquennoy \& Mayor (1991, hereinafter DM91) constrain the period
distribution and the binary proportion for G-type field stars. The
former can be approximated by a Gaussian distribution in log$_{10}a$
with mean near 34~AU, where $a$ is the semi-major axis.  The
proportion of binary systems among G-dwarfs is $f_{\rm G}\approx
0.5\pm0.1$. Virtually the same conclusions hold true for K-~(Mayor et
al. 1992) and M-dwarfs (Fischer \& Marcy 1992) (see also Halbwachs,
these proceedings, and the summary in Kroupa 1995c). Fisher \& Marcy
find that about 7~per cent of all M-dwarfs are triple systems, and
that about 1~per cent are quadruple systems. The proportions are
likely to be higher because the sample of M-dwarf systems within 20~pc
is far from complete. The companion star fraction for M-dwarfs is thus
$f_{\rm comp,M}\approx0.51$ (compare with $f_{\rm M}=0.42\pm0.09$).
Similarly, $f_{\rm comp,G}$ and $f_{\rm comp,K}$ is higher than
$f_{\rm G}$ and $f_{\rm K}$, respectively.  The mass-ratio
distribution for G-dwarf binary systems is biased towards low-mass
companions (Fig.~10 in DM91). The data are still rather poor for
lower-mass primaries but exclude a preponderance of companions with
nearly equal mass (Mayor et al. 1992, Fischer \& Marcy 1992). How
mass-ratio distributions can be miss-interpreted if unknown selection
effects operate is shown by Tout (1991).

That deep star-count data are affected significantly if companion
stars are not counted has been stressed by Buser \& Kaeser (1985),
who construct the system LF from the nearby LF (their Fig.~2) and
apply it to Galactic structure studies. Piskunov \& Malkov (1991)
compute the single-star LF given an observed system LF and assumed
different mass-ratio distributions, as well as a power-law $m(l)$
relation. In their Fig.~1, they show how the system LF significantly
underestimates the stellar number densities. However, the above
studies do not include Malmquist bias nor the method of photometric
distance estimation.

In deep surveys the typical resolution is 2~arc sec (RG), which
implies that multiple stellar systems with dimensions less than
$200\,$AU will not be resolved at a distance of
$100\,$pc. Furthermore, most faint companions will not be seen even if
the system is formally resolved because the photometric distance limit
decreases with decreasing brightness of the star, and because glare
around a bright star can hide faint companions. The great majority of
faint companion stars will be missed.

Counting a multiple star system as a single star has the following
effects: If the companion(s) are bright enough to affect the system
luminosity noticeably, then the estimated photometric distance will be
too small; the remaining companions are lost from the star-count
analysis. The former effect enhances the apparent stellar number
density at brighter magnitudes.  This is countered by the larger
effective photometric distance limit together with the approximately
exponential stellar density fall-off perpendicular to the Galactic
plane.  The second effect reduces the star-counts at faint magnitudes
leading to a significant bias, because a G-, K- and bright M-dwarf
has, on average, a faint M-dwarf companion.

\section{Constraining the MF}
Before proceeding to infer the MF from the LF, we must understand
which information is contained in each of the LFs discussed in
Section~2. Clearly, $\psi_{\rm near}$ is the most direct measurement
of the MF, given $dm/dM_P$.  But $\psi_{\rm near}$ is poorly
constrained for $M_V>12$ by the small number of stars per magnitude
bin, amounting in total to about~20 stars.  The constraints available
so far suggest that $\psi_{\rm near}$ is approximately flat for
$M_V>12$.  The deep surveys show that $\psi_{\rm phot}$ does not vary
with direction, so that they can be combined to improve the
constraints on $\psi_{\rm phot}$. However, not all estimates of
$\psi_{\rm phot}$ have been treated correctly for Malmquist bias,
which depends on the colour--magnitude relation used.  To avoid this
problem, Kroupa (1995a) combined four photometric LFs that base
photometric distance estimation on the $M_V(V-I)$ relation, but are
not corrected for Malmquist bias, and obtained $\overline{\psi}_{\rm
phot}^*$.  Application of Malmquist corrections then gives an estimate
of the distribution of stars with luminosity in the deep surveys,
$\overline{\psi}_{\rm phot}$. This LF implies that 5~stars should have
been found in the sampling volume in which $\psi_{\rm near}$ is
defined. Given that the one-dimensional velocity dispersion of the
stars defining $\psi_{\rm near}$ is about 30~km/sec (KTG93; Reid,
Hawley \& Gizis 1995), it follows immediately that the chance of
observing a local overdensity by a factor of~4 is negligibly small.
It is therefore concluded with high confidence that
$\overline{\psi}_{\rm phot}$ and $\psi_{\rm near}$ do not measure the
same physical quantity. The same mixing argument explains why
$\psi_{\rm phot}$ is independent of direction.

It comes to mind (DLH) that unresolved binary systems are the most
likely cause of the significant underdensity of faint stars in
$\overline{\psi}_{\rm phot}$. This has been verified by
KTG91. Subsequently, Reid (1991) claimed that binary systems cannot
resolve the discrepancy. This, however, is a conclusion drawn upon the
basis of assumptions that define models outside solution space (Kroupa
1995b).

In order to clarify some of the questions posed by KTG91 and Reid
(1991), KTG93 developed a detailed model of star-count data in which
they took into account consistently all the effects discussed above
that influence $\psi_{\rm phot}$. One possible solution is shown in
Fig.~3.
\begin{figure} 
\plotone{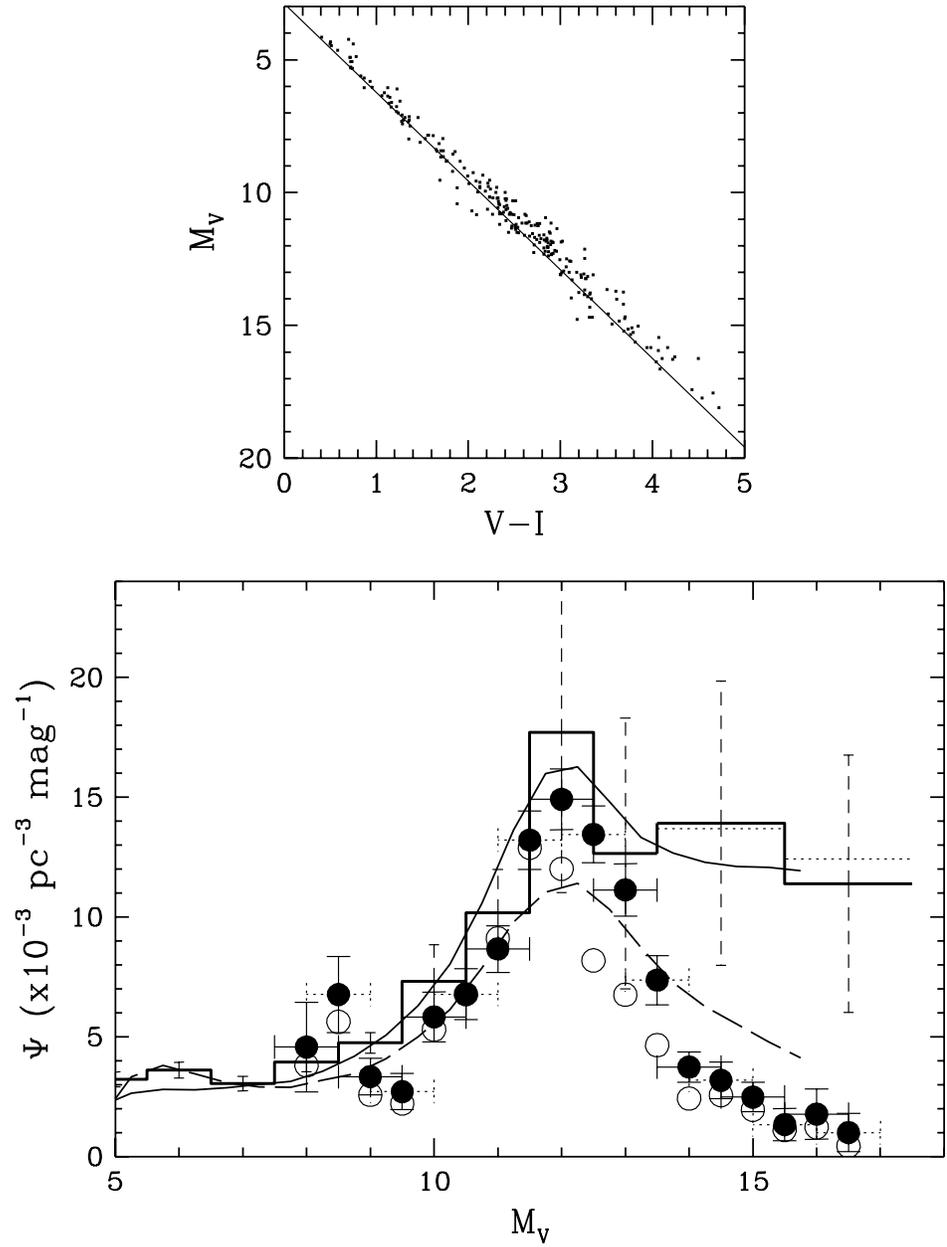}
\caption{Realistic star-count model. See text for details.}
\end{figure}
The upper panel shows the linear $M_V(V-I)$ relation from SIP (solid
line), which is taken to be a model for single stars with the same
solar-neighbourhood metallicity and age. A realistic distribution of
metallicity, age and binarity then leads to a sample of stars that
spread around the underlying relation. The systematic colour-dependent
off set from the underlying relation is evident. This necessitates a
consistent correction of the $M_V(V-I)$ relation fitted to the data. A
complex multi-dimensional Monte-Carlo analysis shows that the SIP star
count data towards the North Galactic Pole can be fitted with the same
MF that also fits $\psi_{\rm near}$.  An exponential Galactic disc,
with $h$ a free parameter, is included explicitly, because unresolved
metal-rich binary systems with nearly equal component masses at a true
distance of 300--400~pc, can enter a photometric distance limit of
about 100~pc (Kroupa 1995b).

The resulting MF is approximated by a two-part power-law below
$1\,M_\odot$.  The IMF for all stellar masses is obtained by extension
with a third power-law segment for more massive stars from the
analysis of Sc:

\begin{displaymath}
\xi_{\rm KTG}\propto m^{-\alpha_i}, 
\end{displaymath}

\noindent 
where $0.70<\alpha_1<1.85$ for $0.08<m/M_\odot\le0.5$, $\alpha_2=2.2$
for $0.5<m/M_\odot\le1$, and $\alpha_3\approx2.7$ for $1<m/M_\odot$
(Fig.~22 and Eqn.~13 in KTG93).  In this notation Salpeter's power-law
index is $\alpha=2.35$.  There is some evidence that $\alpha_3<2.7$
for the IMF (Tsujimoto et al. 1997 and references therein), and for
main-sequence stars in the Galactic field $\alpha_3\approx4.5$.

The model {\it observed} single-star LF becomes the solid curve in the
lower panel of Fig.~3. It is affected significantly by the spread of
metallicities and trigonometric distance uncertainties, and compares
very well with the real $\psi_{\rm near}$ shown as the solid histogram
(from Kroupa 1995a). The thin dotted histogram shows the effect of
counting Gl~866C and shifting the sampling distance from 5.2~pc to
5.23~pc.  The dashed curve is the model photometric LF with Malmquist
bias not removed, assuming a companion star fraction $f=0.7$ and
$h=270$~pc. The solid dots show $\overline{\psi}_{\rm phot}^*$ (binned
into one-magnitude wide bins offset by half a magnitude), which is the
relevant observational constraint, and the open circles are
$\overline{\psi}_{\rm phot}$.

In the above model, stars are combined randomly from the IMF to make
the uncorrelated mass ratio distribution, $f_q$, for binary
systems. This procedure does not give an $f_q$ that is consistent with
the observational constraints (Section~5) because too many low-mass
companions are produced. That the MF derived with the above analysis
remains a good solution if $f_q$ consistent with the observational
constraints is used, is shown in Fig.~4. In the upper left panel
($m_{\rm S}$ and $m_{\rm P}$ are the mass of the secondary and
primary, respectively) the model $f_q$ for G-dwarf primaries is the
solid histogram, which can result from the initial uncorrelated
distribution (dot-dashed histogram) if stars form in embedded clusters
(Kroupa 1995d). The observed data (DM91) are the solid dots. The
overall binary proportion is $f=0.48$. Of these about 9~per cent have
periods shorter than $10^3$~d, with $f_q$ given by the solid histogram
in the upper right panel of Fig.~4. The observed data (Mazeh et
al. 1992) are the solid dots. The model $f_q$ for all primaries less
massive than $1.1\,M_\odot$ and all periods is shown in Fig.~12 in
Kroupa (1995d).  The single star LF for a population with equal metal
abundance, age and no distance errors is the solid curve in the lower
panel of Fig.~4 (compare to Fig.~3). The corresponding system LF is
the dashed curve, and the relevant observational data are the solid
dots ($\overline{\psi}_{\rm phot}$).

\begin{figure} 
\plotone{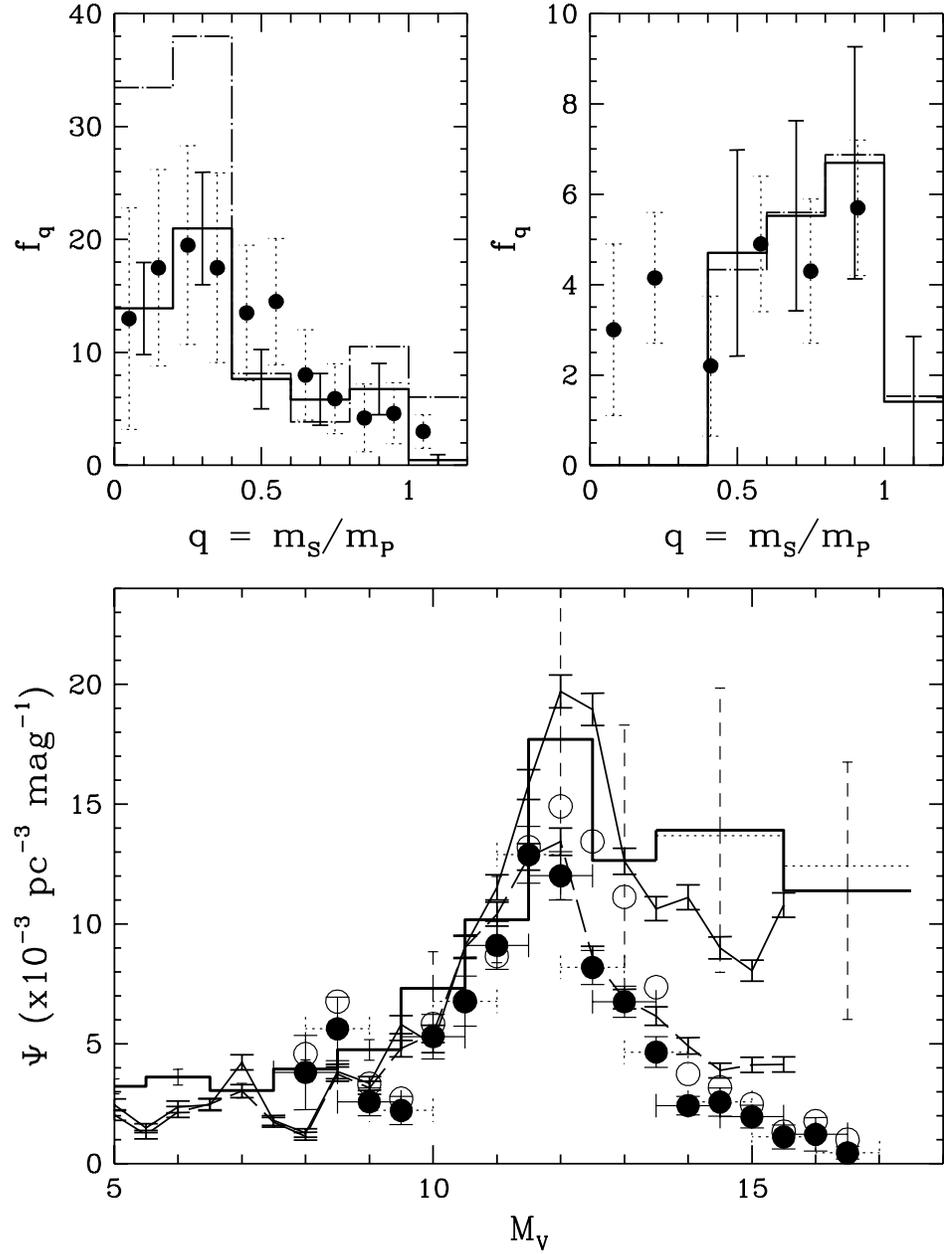}
\caption{Realistic binary star model. See text for details.}
\end{figure}

Comparison of Figs.~3 and~4 shows that the model photometric LFs
contain too many systems for $M_V>14$. As discussed in Kroupa (1995b),
this is not to be taken too seriously because, among other effects,
photometric calibration may be uncertain for the faint red
dwarfs. J. Gizis (private communication) points out that the hitherto
neglected non-linearity of the $M_V(V-I)$ relation (Section~4 in
Leggett, Harris \& Dahn 1994; Fig.~1 in Baraffe et al. 1995) increases
$\psi_{\rm phot}$ at the faint magnitudes, which improves agreement
with the models. The model shown in Fig.~4 does not rely on adoption
of a colour-magnitude relation.

It is thus possible to unify $\psi_{\rm near}$ and $\psi_{\rm phot}$.
The above analysis leads to the most stringent constraints on the MF
available because both the single-star (i.e. $\psi_{\rm near}$) {\it
and} the much better constrained system LF (i.e. $\psi_{\rm phot}$)
are used simultaneously.  Nevertheless, the most direct path to the MF
is via the single-star LF. Only a major, careful and time-consuming
observational effort (e.g. Henry et al. 1997) will improve the poorly
constrained $\psi_{\rm near}$, but it is instructive to consider the
MF that results from the direct conversion (see Section~1) of the
presently available $\psi_{\rm near}$ using different $m(M_P)$
relations: D'AM86 find $1.64<\alpha <2.44$ for $0.1<m/M_\odot<0.3$
using the Wielen et al. (1983) LF and their theoretical
mass--luminosity relation; HM90 find $\alpha \approx0.8$ for
$0.08<m/M_\odot<0.5$ using the nearby LF they estimate in the K-band
and their empirical $m(M_K)$ relation; Laughlin \& Bodenheimer (1993)
find $\alpha\approx 2.35$ for $0.1<m/M_\odot<0.5$ using $\psi_{\rm
near}(M_K)$ from HM90 and their own mass--luminosity relation; Haywood
(1994) finds $1.3<\alpha<2.3$ for $0.12<m/M_\odot<0.35$ using $\psi_{\rm
near}(M_V)$ from DLH and stellar models from VandenBerg; Kroupa
(1995b) finds $0.66<\alpha_1<1.44$ for $\psi_{\rm near}(M_V)$ and
$m(M_V)$ from KTG93; and Mera et al. (1996) find $\alpha \approx
2\pm0.5$ for $m<0.6\,M_\odot$ using $\psi_{\rm near}$ from KTG93 and
their own $m(M_V)$ relation.  Clearly, more or less the same estimate
for the single-star LF, $\psi_{\rm near}$, yields quite different
results depending on which mass--luminosity relation is used. Another
major source of bias in all these estimates of the MF is the neglect
of the metallicity spread and parallax errors. This is demonstrated by
comparing $0.66<\alpha_1<1.44$ (Fig.~3 in Kroupa 1995b) with the model
that does include these effects: $0.6<\alpha_1<2.3$ (Fig.~16 in
KTG93).

An entirely independent estimate of the MF is provided by Tsujimoto et
al. (1997), who find $\alpha>1.8$ down to below the hydrogen burning
mass limit using a model for the chemical enrichment history of the
Galaxy, and the additional constraint given by the mass-to-light ratio
of stars in the solar neighbourhood.  Finally, it is interesting to
note that Zhao, Rich \& Spergel (1996) construct a consistent model of
the Galactic bar, and compare the distribution of microlensing
time-scales with predictions assuming different stellar MFs.  Their
Fig.~3 shows that it is possible to obtain reasonable agreement with
the distribution of time-scales if the MF is flat below $0.5\,M_\odot$
($\alpha=0$; but see Chabrier, these proceedings), and also if it
continues as a Salpeter ($\alpha=2.35$) power-law to lower masses, as
long as there is no significant population of brown dwarfs.

\section{Discussion and Conclusions}
Below approximately $0.5\,M_\odot$ the IMF may be conveniently
approximated by a power law with $\alpha\approx1.5$ down to the
hydrogen burning mass limit (see also Larson 1992).  Whether the MF
continues as some log-normal function to smaller masses
(e.g. Zinnecker 1984; Adams, these proceedings) cannot yet be decided
upon (see also Malkov, Piskunov \& Shpil'kina 1997).

Main-sequence stars ($\alpha_3=4.5$) contribute
$\Sigma_*=30\pm8\,M_\odot$ pc$^{-2}$ to the Oort limit for the
three-part power-law MF with $0.7<\alpha_1<1.85$ and $h\approx
300$~pc. The inter-stellar medium contributes $\Sigma_{\rm
ISM}=13\pm3\,M_\odot$ pc$^{-2}$ (Kuijken \& Gilmore 1989) and stellar
remnants contribute $\Sigma_{\rm rem}=4.5\pm1.5\,M_\odot$ pc$^{-2}$
(Weidemann 1990). It follows that the dynamically inferred
mass-density in the Galactic disc ($\Sigma_*=48\pm9\,M_\odot$
pc$^{-2}$, Kuijken \& Gilmore 1991) can be accounted for. It also
follows that dark matter, e.g. in the form of brown dwarfs, cannot be
a significant mass-component.  The steeper MFs discussed in Section~3
imply essentially the same $\Sigma_*$ (Mera et al. 1996), but it
remains to be seen if such MFs can also fit $\overline{\psi}_{\rm
phot}$. For this, detailed modelling (Section~6) is required. Care
must be taken to ensure that a $m(M_P)$ relation is used that fits the
observational constraints (Fig.~2), that its first derivative is
continuous, and that it is defined on an adequately fine stellar-mass
grid (KT).

LFs extending down to approximately the hydrogen burning mass limit
have been observed for the young Pleiades (Hambly, Hawkins \& Jameson
1991) and the dynamically evolved Hyades (Reid 1993) clusters.  The
power-law MF for Galactic field stars with $\alpha_1=1.3$ fits both
LFs (Kroupa 1995e).  Observed LFs must be corrected consistently for
unresolved binary stars, but care must be taken not to assume the same
binary-star orbital characteristics in open clusters as in the
Galactic field.  The Pleiades and Hyades clusters are not typical
reservoirs of Galactic field stars, so there does not appear to be
much evidence for an environmentally dependent IMF for low-mass stars.

\acknowledgments I am grateful to Michael Hawkins for a very memorable
evening ten years ago at the Siding Spring Observatory, where he gave
me my first introduction to this subject.  I thank C. A. Tout for
proof-reading the manuscript.

\end{document}